\documentstyle[epsf,prb,aps,preprint]{revtex}
\draft
\begin{document} 
\title{\textbf{A Geometric Formulation of Quantum Stress Fields}} 
\author{Christopher L. Rogers and Andrew M. Rappe}
\address{Department of Chemistry and Laboratory for Research 
on the Structure of Matter, \\
University of Pennsylvania, Philadelphia, PA 19104-6323.}  
\date{\today}
\maketitle

\begin{abstract}
We present a derivation of the stress field for an interacting
quantum system within the framework of local density functional theory.
The formulation is geometric in nature and exploits the relationship
between the strain tensor field and Riemannian metric tensor field.
Within this formulation, we demonstrate that the stress field 
is unique up to a single ambiguous parameter.
The ambiguity is due to the non-unique dependence of the kinetic 
energy on the metric tensor.
To illustrate this formalism, 
we compute the pressure field for two phases of solid molecular hydrogen.
Furthermore, we demonstrate that qualitative results obtained by 
interpreting the hydrogen pressure field are not influenced by the presence
of the kinetic ambiguity.
\end{abstract}
\section{Introduction}
The stress, or the energetic response to deformation or strain, plays
an important role in linking the physical properties of a material
(e.g.\ strength, toughness) with the behavior of its microstructure.
In addition, the spatial distribution of stress is an invaluable tool
for continuum modeling of the response of materials. The stress
concept has been applied at atomistic scales as well. Over the last
fifteen years, there has been a continuing trend toward understanding
various structural and quantum-mechanical phenomena in materials in
terms of their response to stress \cite{ibachrev}.

For example, the residual stress at equilibrium has been used to
assess the structural stability of systems containing surfaces or
strained interfaces. It has been demonstrated that the
desire to minimize surface stress can give rise to reconstructions on
high symmetry surfaces
\cite{needs1,needs2,scheffler,ibach,filippeti1,vanderbilt1,vanderbilt2},
and the stability of epitaxially grown bimetallic systems has been
attributed to the formation of incommensurate overlayers, defects, and
dislocations which minimize the stress near the metal-metal interface
\cite{agcu1,agcu2}. The stress can have significant effects on
chemical reactivity as well. It has been shown that small molecule
chemisorption energies and reaction barriers on certain strained metal
and semiconductor surfaces are quite different from those on
the corresponding unstrained surface \cite{vanderbilt3,norskov}.

Formally, studies of the above phenomena must include a
quantum-mechanical description of the system's electronic degrees of
freedom. Therefore, one must consider how a stress is defined quantum
mechanically. Methods for calculating the stress in quantum mechanical
systems have been developed since the birth of quantum theory \cite{schro}. 
However, research in developing a formalism for
determining the quantum stress in solid-state systems has recently
been revitalized. This is mainly due to ever-increasing opportunities
to perform accurate and efficient quantum-mechanical calculations on
systems which exhibit stress mediated phenomena.

The stress is a rank-two tensor quantity, usually taken to be
symmetric and therefore torque-free. Two useful representations of the
stress tensor are the volume-averaged or total stress, $T_{ij}$, 
and the spatially varying stress field $\sigma_{ij}
(\bbox{\mathrm{x}})$. The two representations are related, since
the total stress for a particular region in a system is the stress
field integrated over the volume. Nielsen and Martin developed a
formalism for calculating the total quantum stress in periodic systems
\cite{nielsen}. They define the total stress as the variation of the
total ground-state energy with respect to a uniform scaling of the
entire system. This uniform scaling corresponds to a homogeneous or
averaged strain over the entire system. They further demonstrate that
the total quantum stress is a unique and well-defined physical
quantity. Their formulation has been successfully implemented to
study a variety of solid state systems \cite{needs1,needs2}. Other
formalisms for determining the total quantum stress have been created
as well \cite{needs1,passerone,resta}.

Although these formalisms have provided important tools for studying
quantum stress, the stress field is a more useful quantity that
contains important information regarding the distribution of the
stress throughout the system. A knowledge of the spatial dependence
of the quantum stress is vital if one wishes to predict the spatial
extent of structural modifications or understand phenomena at
interfaces in complex heterogeneous systems. However, 
the quantum stress field can only be specified up to a so-called gauge term. 
Therefore the quantum stress field is not unique.
A traditional way to develop a quantum stress field formalism is to
consider the stress field's relationship with the force field. From
this perspective, the stress field can be defined as any rank-two
tensor field whose divergence is the force field of the system:
\begin{eqnarray}
F^{i}(\bbox{\mathrm{x}})= \nabla_{j} \sigma^{ij}(\bbox{\mathrm{x}}). \label{force-stress}
\end{eqnarray}
(Note that the Einstein summation convention for repeated indices is used throughout the paper.)
One can add to $\sigma^{ij}$ a term of the form
\begin{eqnarray}
\frac{\partial }{\partial x^{k}}A^{ijk}
(\bbox{\mathrm{x}}),
\end{eqnarray}
where $A^{ijk}$ is any tensor field antisymmetric in
$j$ and $k$, and recover the same force field, thereby
demonstrating the non-uniqueness of this stress field definition. General
formulations for computing stress fields in quantum many-body systems 
have been derived by Nielsen and Martin, Folland, Ziesche and co-workers, and Godfrey
\cite{nielsen,folland,ziesche,godfrey}. Regardless of issues concerning non-uniqueness,
the quantum stress field has been used to investigate material properties.
For example, Ramer and co-workers developed a method to calculate the 
resultant stress field from an induced homogeneous strain in order
to study piezoelectric effects in perovskites \cite{ramer}.
They incorporate the additional constraint that the field must be the smoothest fit to the
ionic forces. However this method cannot be used to calculate the residual
stress field at equilibrium, nor can it determine the energy
dependence on strains which do not have the periodicity of the unit
cell. Filippetti and Fiorentini developed a formulation of the stress
field based on the energy density formalism of Chetty and Martin
\cite{filippetti2,chetty}. Since this formulation is based explicitly
on the energy density, which cannot be uniquely specified, their stress field
is also ambiguous.

In order to ascertain the exact nature of these ambiguities and
to determine their effect on using the quantum stress field
as an interpretive tool, we have developed a 
formulation of the quantum stress field based on differential
geometry. This approach is similar to  Mistura's work in deriving
pressure fields for classical inhomogeneous fluids \cite{mistura}.
We first develop the necessary geometric framework and discuss
the origin of ambiguities in the stress. We then derive
the stress field in the context of local density functional
theory (LDA-DFT). Finally, we apply our formulation by
using the pressure field to explain the energetic
ordering between two phases of solid molecular hydrogen.

\section{Metric and Strain Fields}
The geometric formulation for the stress field is developed from the
relationship between a strain tensor field and the Riemannian metric
tensor field. This relationship has been derived for flat manifolds
\cite{fung} and for manifolds of arbitrary curvature \cite{mistura}.
The infinitesimal squared distance between two points on a Riemannian
manifold can be written as $ds^2 = g_{ij} dx^i dx^j,$ where the metric
$g_{ij}$ is a rank two symmetric tensor field.  The system can then be
infinitesimally distorted so that the positions of the points change,
$dx'^i = dx^i + du^i,$ which changes the squared distance to $ds'^2 =
g_{ij} dx'^i dx'^j$.  However, a coordinate transformation can be made
so that the new distance can be computed from the point of view of the
predistorted positions: $ds'^2 = g'_{ij} dx^i dx^j$, with the new
metric defined as
\begin{eqnarray}
g'_{kl} = g_{ij} \left( \delta^{i}_{k} + \partial_{k} u^i \right) \left( \delta^{j}_{l} + \partial_{l} u^j \right).
\end{eqnarray}
The strain field $\epsilon_{ij}$ can be related to the difference between distances before and after a distortion has been applied:
$ds'^2 - ds^2 = 2 \epsilon_{ij} dx^i dx^j$. Therefore the metric $g'_{ij}$ can be related to the strain:
\begin{eqnarray}
g'_{ij} = 2\epsilon_{ij} + g_{ij}. \label{strain-metric}
\end{eqnarray}

We define the stress field $\sigma^{ij}$ via the virtual work relationship between energy and strain
\begin{eqnarray}
\delta E = \int \sqrt{g} \sigma^{ij} \delta \epsilon_{ij} d^3x, \label{vwork}
\end{eqnarray}
where $g$ is the determinant of $g_{ij}$.
Using Eq.\ \ref{strain-metric}, the stress field can be expressed as
\begin{eqnarray}
\sigma^{ij} &=& \frac{1}{\sqrt{g}} \frac{\delta E}{\delta \epsilon_{ij}} = \frac{2}{\sqrt{g}} \frac{\delta E}{\delta g_{ij}} \nonumber \\
\sigma_{ij} &=& - \frac{2}{\sqrt{g}} \frac{\delta E}{\delta g^{ij}}, \label{stress-metric}
\end{eqnarray}
where $g^{ij}$ is the inverse of the metric. It is this definition which we will use to compute the stress at the
quantum length scale, with $E=\langle \Psi |\hat{H}| \Psi \rangle$ defined as the energy of the ground-state $|\Psi \rangle$ 
of a system described by a Hamiltonian operator $\hat{H}$. 
This method of deriving the stress field is identical to that used to compute the ``improved'' energy-momentum tensor in field theories which are
coupled to a background gravitational source \cite{landau2}.
Specifically, we shall be interested in the value of
the stress field evaluated at the Euclidean metric ($g_{ij} = \delta_{ij}$). 

\section{Uniqueness}
For the virtual work theorem, (and therefore Eq.\ \ref{stress-metric}), to specify a unique stress field, 
two criteria must be obeyed. First, the metric must be varied freely, without constraints, when computing the
functional derivative. Second the quantum-mechanical total energy must be unambiguously defined for all metrics.
Before we can explain the origin and the consequences of these criteria,
we first must discuss in more detail the relationship between a metric tensor
and the geometry of the underlying space containing the physical system.
Metric tensor fields, for our purposes, can be classified into two categories:
Those fields which cause the Riemann curvature tensor to vanish at every point in space
(e.g. the Euclidean metric $g_{ij}=\delta_{ij}$), and those fields which give rise to
a non-zero Riemann curvature tensor (e.g. the metric for the two-dimensional surface of a sphere).
The latter class of metrics will be referred to as non-Euclidean.
The Riemann curvature is a rank four tensor field defined as
\begin{eqnarray}
R_{iklm} &=& \frac{1}{2} \left( \partial_{k} \partial_{l} g_{im} + \partial_{i} \partial_{m} g_{kl}
- \partial_{k} \partial_{m} g_{il} - \partial_{i} \partial_{l} g_{km} \right)
+ g_{np} \left( \Gamma^{n}_{kl} \Gamma^{p}_{im} - \Gamma^{n}_{km}\Gamma^{p}_{il} \right),
\end{eqnarray}
where the matrices $\Gamma^{i}_{kl}$ are the Christoffel symbols:  
\begin{eqnarray}
\Gamma^{i}_{kl} &=& \frac{1}{2} g^{im} \left(\partial_{l}g_{mk} + \partial_{k}g_{ml} - \partial_{m}g_{kl} \right).
\end{eqnarray}
Metrics which give zero Riemann curvature over the entire manifold are said to be flat. A coordinate transformation
can always be performed so that any flat metric will transform into the Euclidean metric over the entire manifold.

We can think of the system as a collection of particles and/or fields
embedded in a manifold $M$ diffeomorphic to $R^3$. We will refer to $M$ as the background manifold. 
$M$ is a flat space and therefore the Euclidean metric can be used
to compute distances over the entire manifold. Obviously any physically possible rearrangement is constrained
to keep the particles embedded in $M$ (assuming no gravitational field is present). 
Therefore the system should only be able to realize  
strain configurations which correspond (via the relation in Eq.\ \ref{strain-metric}) to metric tensor fields
which do not produce any Riemann curvature, since $M$ is flat. This implies that the
virtual work relation in Eq.\ \ref{vwork} should be rewritten to reflect this constraint.
All flat metrics can be expressed as
\begin{eqnarray}
g_{ij} &=& \delta_{ij} + \partial_{j}u_{i} + \partial_{i}u_{j}, \label{flatmetric}
\end{eqnarray}
where $u_{i}$ is a vector field continuous over the entire manifold. The constrained virtual work relation is then
\begin{eqnarray}
\delta E &=& \int \sqrt{g} \sigma^{ij} \left( \partial_{j} \delta u_{i} + \partial_{i} \delta u_{j} \right) d^3x.
\end{eqnarray}   
Since we are now constrained to a flat space we can choose Cartesian coordinates so that $\sqrt{g}=1$:
\begin{eqnarray}
\delta E &=& \int \sigma^{ij} \left( \partial_{j} \delta u_{i} + \partial_{i} \delta u_{j} \right) d^3x \\
         &=& - \int \left( \partial_{j} \sigma^{ij} \delta u_{i} + \partial_{i} \sigma^{ij} \delta u_{j} \right). \label{newvwork}
\end{eqnarray}
Notice that the divergence of the stress field has appeared in Eq.\ \ref{newvwork}, so this approach cannot be used
to uniquely specify the stress field. Therefore introducing a constraint on variations in the metric, has
produced an ambiguity in the stress field. This is the origin of the aforementioned first criterion for
uniqueness of the stress field. To have a unique stress field, the system must be allowed to realize the 
energetics of \textit{any} applied strain configuration which can be represented by a continuous symmetric tensor field.

If we hypothesize that stress fields are unique, then we must explain physically how the system can be 
altered so that the Riemann curvature at some point is non-zero. 
Even a first order change away from the Euclidean metric will generate a first order change in the curvature.
One way which such distortions could be realized is by introducing a gravitational ``point-like'' 
source. Then, according to General Relativity,
the system would be altered by the fact that the source has generated curvature within the background manifold.
It is through this thought-experiment that the link between the definition of the stress field used in this work
and the energy-momentum tensor in the gravitational literature is established. Unfortunately this is not a
phenomenon that can be observed by experiment, especially on the quantum scale. 

In addition to the gravitational interpretation, we can also consider the meaning of non-Euclidean metrics
within the context of the continuum theory of solids. In this theory, the solid within the limit of vanishing lattice parameter 
can be described by a vector field $u_{i}$ which indicates the displacement of atomic positions away from some reference 
(usually taken to be a perfect crystal) \cite{landau}. If the specified atomic configuration contains 
defects such as disclinations and dislocations, then the displacement field $u_{i}$
will not be perfectly smooth and therefore will violate the integrability conditions:
\begin{eqnarray}
\left( \partial_{i}\partial_{j} - \partial_{j}\partial_{i} \right) u_{k} &=& 0. \label{inte} 
\end{eqnarray}
Within linear elasticity theory, the information regarding the disclinations and dislocations
can be represented by a symmetric tensor field known as the defect density \cite{kleinert}:
\begin{eqnarray}
\eta_{ij} &=& \varepsilon_{i}^{kl}\varepsilon_{j}^{mn}\partial_{k}\partial_{m}u_{ln},
\end{eqnarray}
where $\varepsilon_{ikl}$ is the Levi-Civita symbol, and $u_{ij}$ is the continuum strain tensor
field. Notice that $\eta_{ij}$ is the double curl of the strain field. If $\eta_{ij}=0$ 
over the entire manifold, then the strain field may be written in terms of the gradient of a displacement vector field
$u_{i}$ which obeys the condition in Eq.\ \ref{inte}. Specifically,
\begin{eqnarray}
u_{ij} &=& \frac{1}{2} \left( \partial_{i} u_{j} + \partial_{j} u_{i} \right),
\end{eqnarray}
This is known as compatibility. If there are no defects present, then
the strain field is compatible with a smooth vector displacement field.

The defect density has a rigorous geometric meaning. It can be shown that \cite{kleinert}
\begin{eqnarray}
\eta_{ij}= R_{ij} - \frac{1}{2}g_{ij}R = G_{ij}.
\end{eqnarray}
$G_{ij}$ is known as the Einstein curvature tensor and is related to the Riemann curvature through the Ricci curvature
$R_{ij} = g^{lm}R_{limj}$ and the scalar curvature $R = g^{ij}R_{ij}$.
Therefore if a particular strain configuration in a continuum model generates Einstein curvature,
it should be interpreted as a mathematical representation of the discontinuity in the displacement
of the particles in the system due to defects. The curvature
should not be thought of as a physical manipulation of the geometry of the background manifold.
However, on the quantum scale, the relationship between the curvature and the compatibility of the strain field
requires careful consideration. It is not obvious how to assign physical meaning to, for example, 
a ``defect'' that is point-like in an electron gas.

Since we must consider the energetic response of the system to non-Euclidean metrics, the quantum-mechanical energy
must be generalized so that it is unambiguously defined for metrics that generate an arbitrary Riemann curvature. 
This is the origin of the second criterion for uniqueness of the quantum stress field. 
If we take the \textit{ab-initio} perspective, we can build the generalized Hamiltonian from only a few facts.
First it is reasonable to require that the energy transform as a covariant scalar regardless of
the geometry. Second, the generalized Hamiltonian must reduce to the known correct flat-space result when evaluated with a metric 
which generates zero Riemann curvature. Finally, the Hamiltonian must be well-behaved in the sense that
it corresponds to a quantum field theory that is renormalizable. Unfortunately, these conditions are not strict enough to uniquely specify a single Hamiltonian.
There are an infinite number of possible choices, all which satisfy the above three conditions.
In particular, the ambiguity manifests itself in the kinetic energy.
In Euclidean space, the Hamiltonian for a spin-zero free field $\hat{\psi}$ is
\begin{eqnarray}
\hat{H}  &=& \frac{1}{2} \int \nabla \hat{\psi}^{\dagger} \cdot \nabla \hat{\psi} d^3x.
\end{eqnarray}
To generalize for a manifold of arbitrary curvature, we can write $\hat{H}$ in a covariant form:
\begin{eqnarray}
\hat{H}'  &=& \frac{1}{2} \int \sqrt{g} g^{ij} \partial_{i} \hat{\psi}^{\dagger} \partial_{j} \hat{\psi} d^3x .
\end{eqnarray}
The operator $\hat{H}'$ satisfies the aforementioned requirements for a generalized Hamiltonian. 
Notice $\hat{H}'$ could have been constructed by writing $\hat{H}$ in terms of curvilinear coordinates in a flat space.
This is known as the principle of least coupling. It is the simplest way to generate operators which
are suitably generalized for curved spaces. Also it is the mathematical formulation of the equivalence principle
in General Relativity. However, we could postulate another generalized form of $\hat{H}$:
\begin{eqnarray}
\hat{H}^{\prime \prime} &=& \int \sqrt{g} \left(\frac{1}{2} g^{ij} \partial_{i} \hat{\psi}^{\dagger} \partial_{j} \hat{\psi}
+ \alpha R  + \beta R \hat{\psi}^{\dagger} \hat{\psi} + 
\gamma R^{ij} \partial_{i} \hat{\psi}^{\dagger} \partial_{j} \hat{\psi} \right) d^3x,  
\end{eqnarray}
where $\alpha, \beta, \gamma$ are coupling constants to the various curvature terms. Note that
$H^{\prime \prime}$ is a scalar and gives the correct Euclidean space Hamiltonian when $g_{ij}=\delta_{ij}$.
However the term multiplied by $\gamma$ will cause the theory to be non-renormalizable in a curved space. This can be determined via dimensional analysis.
In the units $\hbar=1$, $c=1$ (commonly used in relativistic field theories \cite{units}) 
the Hamiltonian must have units of mass, or equivalently (length)$^{-1}$. This
implies the integrand, or Hamiltonian density, must have units of (mass)$^{4}$ since the
measure $d^3x$ has units of (mass)$^{-3}$. (The metric is unitless.)
The spin-zero field $\hat{\psi}$ has units of (mass)$^{1}$,
which can be deduced from the first term of $H''$ which contains only gradients.
The scalar curvature $R$ has units of (mass)$^{2}$, since it contains the second derivatives
and products of the first derivatives of the metric. Hence we can conclude that 
$\alpha$ has units of (mass)$^{2}$, $\beta$ is dimensionless, and $\gamma$ has units
of (mass)$^{-2}$. If a coupling constant has negative mass dimension, then the theory 
is non-renormalizable \cite{renorm}. Therefore we can set $\gamma$ equal to zero.

The ambiguity can be further reduced if we physically reason that if no matter is present in the system, then
the stress field should be zero everywhere i.e. the part of $H''$ multiplied by $\alpha$
is a background term. Furthermore, this term's contribution to the stress field will vanish
when the stress field is evaluated at the Euclidean metric. Consequently we may set $\alpha =0$. 
However, we currently cannot determine any restrictions on the value of $\beta$ from first principles.
This is the origin of the ambiguity in the dependence of the 
energy on the metric. Therefore the quantum stress field will only be unique up to the determination 
of $\beta$. We further note that this problem is well-known in the gravitational and constrained-dynamics
literature, and to date remains unresolved \cite{birrell,saa,baleanu}.

\section{The DFT Stress Field}
We now use the geometric approach to derive the quantum stress field of a many-electron system in
the presence of a fixed set of classical positive charged ions using
local density functional theory \cite{HK,KS}.  The ground state
electronic charge density of the system is written as
$n(\bbox{\mathrm{x}})= \sum_\mu
\phi^{\ast}_{\mu}(\bbox{\mathrm{x}})\phi_{\mu}(\bbox{\mathrm{x}})$, where
$\phi_\mu$ are single-particle orthonormal wavefunctions. 
For this derivation, we assume orbitals with fixed integer occupation
numbers. The extension to metals with Fermi fillings is straightforward, 
simply necessitating use of the Mermin functional instead of the total energy \cite{mermin}.
The total charge density of the system can be written as a sum over all
ionic charges and $n$:
\begin{eqnarray}
\rho(\bbox{\mathrm{x}}) = \sum_\mu 
\frac{Z_\mu}{\sqrt{g}}\delta(\bbox{\mathrm{x}} - \bbox{\mathrm{R}}_\mu) -
n(\bbox{\mathrm{x}}),
\end{eqnarray}
where $Z_\mu$ is the charge of the $\mu$-th ion located at position
$\bbox{\mathrm{R}}_\mu$, and the presence of $\sqrt{g}$ insures proper
normalization of the delta function.  The energy of the system can be
written as the following constrained functional which is appropriately
generalized for arbitrary Riemannian spaces:
\begin{equation}
E = E_{\mathrm{k}} + E_{\mathrm{Coulomb}} +
E_{\mathrm{xc}} - \sum_{\mu} \lambda_{\mu} \left( \int \sqrt{g}
\phi^{\ast}_{\mu} \phi_{\mu} d^3x - 1 \right). \label{totale}
\end{equation}
Here $E_{\mathrm{k}}$ is the single particle kinetic energy including the
aformentioned ambiguity,
$E_{\mathrm{Coulomb}}$ is the classical Coulomb interaction between
the total charge density and itself, and $E_{\mathrm{xc}}$ is the
exchange-correlation energy of the electrons. The appearance of the
last term in Eq.\ \ref{totale} is due to the orthonormality constraint
of the orbitals. (We choose a unitary transformation on $\left
\{\phi_{\mu} \right \}$ which enforces orthogonality.) One can express
$E$ as an integral over an energy density \cite{chetty,burke}.
For convenience, we express the energy terms in Eq.\ \ref{totale} as the following:
\begin{eqnarray}
E_{\mathrm{k}} &=& \int \sqrt{g} \left( \frac{1}{2} \sum_\mu g^{ij}
\partial_{i} \phi^{\ast}_{\mu} \partial_{j} \phi_{\mu} + \beta R n \right) d^3x
\nonumber \\ E_{\mathrm{Coulomb}} &=& \int \sqrt{g} \left( \rho V -
\frac{1}{8\pi} g^{ij} {\mathcal{F}}_{i}
{\mathcal{F}}_{j} \right) d^3x \nonumber \\ E_{\mathrm{xc}} &=&
\int \sqrt{g} \ n \varepsilon_{\mathrm{LDA}}(n) d^3x,
\end{eqnarray}
where ${\mathcal{F}}_{i} = - \partial_{i}V $ is the electric
field due to the Coulomb potential $V$ generated by $\rho$,
$\varepsilon_{\mathrm{LDA}}(n)$ is the LDA exchange-correlation
energy density, and $R$ is the scalar curvature.
To obtain the electronic ground-state energy, we require $\delta E /
\delta \phi^{\ast}_\mu = 0$ with the additional constraints of a fixed metric
($\delta g_{ij} =0$) and a fixed ionic charge density
($\delta \rho = -\delta n $). This implies that the orbitals must obey
the Euler-Lagrange equations
\begin{equation}
-\frac{1}{2\sqrt{g}} \partial_{i} \left(\sqrt{g} g^{ij}
\partial_{j} \phi_\mu \right) + \beta  R \phi_{\mu} + \frac{\delta
E_{\mathrm{Coulomb}}}{\delta n} \phi_\mu + \frac{\delta
E_{\mathrm{xc}}}{\delta n} \phi_\mu = \lambda_\mu \phi_\mu, \label{E-L}
\end{equation}
which can be considered the Kohn-Sham equations for a general Riemannian
manifold. Also, a least-action principle for $E_{\mathrm{Coulomb}}$ 
requires that $\rho$ and $V$ obey the Poisson equation:
\begin{equation}
\frac{1}{\sqrt{g}} \partial_{i}\left(\sqrt{g} g^{ij}
\partial_{j}V \right) = -4\pi \rho. \label{Poisson}
\end{equation}

We now vary the total energy with respect to the
metric. It can be proven that we do not need to consider
variations in the electronic wavefunctions, charge density and potentials, since all
such variations would vanish due to Eq.\ \ref{E-L} and Eq.\
\ref{Poisson}. This is the same principle used in the derivation of
the Hellmann-Feynman force theorem and the
energy-momentum tensor (the variation of the action with respect to metric) in general relativity
\cite{landau2,hellmann,feynman}.   
Performing the variation of the total energy with respect to the metric gives the stress field in 
local density functional theory as:
\begin{eqnarray}
\sigma_{ij} &=& - \sum_\mu \partial_{i}
\phi^{\ast}_{\mu} \partial_{j} \phi_{\mu} + A_{ij} + \frac{1}{4\pi}
{\mathcal{F}}_{i}{\mathcal{F}}_{j} + g_{ij} \left(
\frac{1}{2} \sum_\mu \partial_{k}\phi^{\ast}_{\mu} \partial^{k}\phi_{\mu} \nonumber \right.  \\
& & \left. \mbox{} -\frac{1}{8 \pi} {\mathcal{F}}_{k} {\mathcal{F}}^{k} + n
\varepsilon_{\mathrm{LDA}}(n) - \sum_\mu \phi^{\ast}_{\mu} \phi_{\mu} \left[
\lambda_\mu + V \right] \right) , \label{stressfield}
\end{eqnarray}
where we have used the relation $ \partial \sqrt{g} /\partial
g^{ij} = -\frac{1}{2}\sqrt{g}g_{ij}$. For the moment, we simply denote
the contribution of the kinetic ambiguity to the stress field as $A_{ij}$. 
Using Eq.\ \ref{E-L}, we can rewrite Eq.\ \ref{stressfield} as
\begin{eqnarray}
\sigma_{ij} &=& - \sum_\mu \partial_{i}
\phi^{\ast}_{\mu} \partial_{j} \phi_{\mu} + A_{ij} + \frac{1}{4\pi}
{\mathcal{F}}_{i}{\mathcal{F}}_{j} + g_{ij} \left(
\frac{1}{2} \sum_\mu \partial_{k}\phi^{\ast}_{\mu} \partial^{k}\phi_{\mu} \nonumber \right.  \\
& & \left. \mbox{} + \frac{1}{2\sqrt{g}} \sum_\mu \phi^{\ast}_{\mu} \partial_{\kappa}
\left( \sqrt{g} g^{\kappa k} \partial_{k} \phi_{\mu} \right)  - \beta R n 
- \frac{1}{8 \pi} {\mathcal{F}}_{k} {\mathcal{F}}^{k} + n \left(
\varepsilon_{\mathrm{LDA}}(n) - \frac{\delta E_{\mathrm{xc}}}{\delta n} \right) 
\right) . \label{stressfield2}
\end{eqnarray}
Eq.\ \ref{stressfield2} is most useful when evaluated at the Euclidean metric. 
The $\left \{\phi_\mu \right \}$ are then
solutions to the standard Kohn-Sham equations. From here on, we
will refer to $\sigma_{ij}$ with an implied evaluation at
the Euclidean metric. 

At this point we briefly describe how to obtain the explicit form of $A_{ij}$ in Euclidean space.
We need to evaluate
\begin{eqnarray}
A_{ij}(\bbox{\mathrm{y}}) = -\frac{2\beta}{\sqrt{g(\bbox{\mathrm{y}})}} \int n(\bbox{\mathrm{x}}) \frac{\delta 
\left(\sqrt{g(\bbox{\mathrm{x}})} 
R(\bbox{\mathrm{x}}) \right)}{\delta g^{ij}(\bbox{\mathrm{y}}) } d^3x.
\end{eqnarray}
(Variations in $n$ are zero due to the aforementioned use of the variational principle.)
Keeping in mind we are only interested in the result evaluated for
$g_{ij} = \delta_{ij}$, and $R_{ij}=0$, the only non-zero term is
\begin{eqnarray}
A_{rs}(\bbox{\mathrm{y}}) = -\frac{2\beta}{\sqrt{g(\bbox{\mathrm{y}})}}  \int \sqrt{g(\bbox{\mathrm{x}})} 
n(\bbox{\mathrm{x}}) g^{ik}(\bbox{\mathrm{x}}) \frac{\delta 
R_{ik}(\bbox{\mathrm{x}})}{\delta g^{rs}(\bbox{\mathrm{y}}) } d^3x.
\end{eqnarray}
After substantial manipulation and an integration by parts, we obtain
\begin{eqnarray}
A_{rs}(\bbox{\mathrm{y}}) = \frac{2\beta}{\sqrt{g(\bbox{\mathrm{y}})}} \int \sqrt{g(\bbox{\mathrm{x}})} 
\partial_{l} n(\bbox{\mathrm{x}}) 
\left( g^{ik}(\bbox{\mathrm{x}}) \frac{\delta \Gamma^{l}_{ik}(\bbox{\mathrm{x}})}{\delta g^{rs}(\bbox{\mathrm{y}}) } 
- g^{il}(\bbox{\mathrm{x}}) \frac{\delta \Gamma^{k}_{ik}(\bbox{\mathrm{x}})}{\delta g^{rs}(\bbox{\mathrm{y}}) } \right)d^3x,
\end{eqnarray}
plus additional terms which are zero when $g_{ij} = \delta_{ij}$.
Explicit computation of the functional derivatives gives the final result for Euclidean space:
\begin{eqnarray}
A_{ij} = \beta \left( \partial_{i}\partial_{j}n - \delta_{ij} \partial^{k}\partial_{k} n \right).
\end{eqnarray}

Therefore, the complete stress field evaluated at $g_{ij} = \delta_{ij}$ is:
\begin{eqnarray}
\sigma_{ij} &=& - \sum_\mu \partial_{i}
\phi^{\ast}_{\mu} \partial_{j} \phi_{\mu} + \beta \left( \partial_{i}\partial_{j}n - \delta_{ij} \partial^{k}\partial_{k} 
n \right) + \frac{1}{4\pi}
{\mathcal{F}}_{i}{\mathcal{F}}_{j} + \delta_{ij} \left(
\frac{1}{2} \sum_\mu \partial^{k}\phi^{\ast}_{\mu} \partial_{k}\phi_{\mu} \nonumber \right.  \\
& & \left. \mbox{} + \frac{1}{2} \sum_\mu \phi^{\ast}_{\mu} \partial^{k}
\partial_{k} \phi_{\mu} 
- \frac{1}{8 \pi} {\mathcal{F}}^{k} {\mathcal{F}}_{k} + n \left(
\varepsilon_{\mathrm{LDA}}(n) - \frac{\delta E_{\mathrm{xc}}}{\delta n} \right) 
\right) . 
\end{eqnarray}

It is important to note several key features of the form of
$\sigma_{ij}$. First the ambiguity in the kinetic stress is identical to 
that identified by Godfrey via a different formalism\cite{godfrey}.
Also, the kinetic stress contains diagonal terms which are
similar to the symmetric and antisymmetric kinetic energy densities.
The contribution of the exchange-correlation energy to
the stress field is only in the diagonal (pressure-like) terms, which
is the proper behavior for local density functionals \cite{nielsen}
and is identical to the exchange-correlation stress derived in Ref. 21.

The Coulombic contribution to $\sigma_{ij}$ is equivalent to the 
classical Maxwell stress field. This Coulombic term can be obtained by 
Filippetti and Fiorentini's formalism if one chooses 
the so-called Maxwell gauge \cite{filippetti2}. We emphasize that it has been mathematically
proven in the gravitational literature that the usual energy-momentum tensor for electromagnetism 
is unique \cite{collinson,lovelock,kerrighan}. Since this energy-momentum tensor is equivalent to the Maxwell stress in 
the non-relativistic limit and the absence of magnetic fields, 
the classical Coulombic interaction in the stress field is rigorously free 
of ambiguities. 

By integrating the stress field over all space, note that we can obtain the 
total stress $T_{ij}$:
\begin{eqnarray}
T_{ij} &=& \int \left \{ - \sum_{\mu} \partial_{i}
\phi^{\ast}_{\mu} \partial_{j} \phi_{\mu} + \frac{1}{4\pi}
{\mathcal{F}}_{i}{\mathcal{F}}_{j} - \delta_{i j}
\frac{1}{8\pi} {\mathcal{F}}_{\gamma}{\mathcal{F}}^{\gamma} \nonumber \right. \\
& & \left. \mbox{} + \delta_{ij} n \left(\varepsilon_{\mathrm{LDA}}(n) -
\frac{\delta E_{\mathrm{xc}}}{\delta n} \right) \right \} d^3x,
\end{eqnarray} 
which is identical to the expression derived by Nielsen and
Martin\cite{nielsen}. 

It is also instructive to consider the form of the pressure field $P$ defined
as one-third the trace of the stress field. Using our formulation for $\sigma_{ij}$
we obtain:
\begin{eqnarray}
P = \frac{1}{3} \left( \frac{1}{2} \sum_{\mu} \partial^{k}\phi^{\ast}_{\mu} \partial_{k}\phi_{\mu}
- 2\beta \partial^{k}\partial_{k}n - \frac{1}{8\pi} {\mathcal{F}}^{k}{\mathcal{F}}_{k} \right)
+ \frac{1}{2} \sum_{\mu} \phi^{\ast}_{\mu} \partial^{k}\partial_{k} \phi_{\mu} + 
n \left(\varepsilon_{\mathrm{LDA}}(n) - \frac{\delta E_{\mathrm{xc}}}{\delta n} \right). \label{pfield}
\end{eqnarray}
Notice the pressure field contains the Maxwell energy density 
and a combination of the symmetric and antisymmetric kinetic energy densities.
The contribution of the LDA exchange-correlation pressure is consistent
with the unambiguous exchange-correlation energy density derived by Burke and co-workers
\cite{burke2}.

Also present is the undetermined contribution of the Laplacian of the charge density.
We should consider whether this kinetic ambiguity impairs us from using the
the pressure field to obtain novel qualitative information from total energy calculations.
We would therefore need to determine a realistic range for the parameter $\beta$.
Such a range can be derived if we consider the various forms of the kinetic energy
density. Symmetric and antisymmetric forms can be converted from one to the other
by adding a term proportional to the Laplacian of the charge density: 
\begin{eqnarray}
\frac{1}{2} \sum_{\mu} \partial^{k}\phi^{\ast}_{\mu} \partial_{k}\phi_{\mu} - \frac{1}{4} 
\partial^{k}\partial_{k} n = - \frac{1}{4}\sum_{\mu} \left(
\phi^{\ast}_{\mu} \partial^{k}\partial_{k} \phi_{\mu} + \phi_{\mu} \partial^{k}\partial_{k} \phi^{\ast}_{\mu} \right).
\end{eqnarray}
Hence a realistic range for $\beta$ in Eq.\ \ref{pfield} would be from 0 to 1/8.

\section{Application: Solid Hydrogen}
In order to demonstrate the utility of our stress field formalism,
we compute within DFT the pressure field, for two phases of solid molecular hydrogen 
under external hydrostatic pressure of 50 GPa. \cite{calc1}.
Both structures consist of stacked two-dimensional triangular
lattices of hydrogen molecules, with the molecular axis parallel to
the stacking direction and a repeat unit of two layers.  The \textit{m}-hcp
structure has alternating layers shifted so that each hydrogen
molecule is directly above triangular hollow sites in the neighboring
layers.  The second structure (belonging to the \textit{Cmca} space group) has
a different shift, so that each molecule lies directly above midpoints
between nearest-neighbor pairs of molecules in adjacent layers \cite{solidh1}.
The energetics and electronic properties of both structures have been studied 
extensively from first principles \cite{solidh2,solidh3}.

Examination of the pressure field permits us to rationalize the energy
ordering of these structures. The \textit{m}-hcp structure is
energetically favored by 60 meV/molecule.  Figure \ref{fig1}A shows a
contour plot of the pressure field for the \textit{Cmca}
structure. Here we have set the constant $\beta$
equal to zero. The pressure is tensile (greater than zero) through the
interstitial region, indicating that contraction is locally favorable.
The tensile pressure is greatest in the volume directly above and below each
molecule, averaging 3 eV/\AA$^3$. This implies that the system would
energetically favor increased intermolecular coordination. In Figure
\ref{fig1}B we show a similar plot for the \textit{m}-hcp
structure. Again the pressure within the interstitial region is
tensile. However, the pressure field has significantly rearranged, and
the pressures in the regions above and below molecules have been
reduced to approximately 2.25 eV/\AA$^3$.  It is also clear from the
charge density plots (Figure \ref{fig1}C and \ref{fig1}D) that the
reduction in pressure is correlated with an increase in bonding
between molecules and increased charge delocalization.  The pressure
fields within the molecules are compressive, and they are several
orders of magnitude larger than the interstitial features.  However,
because these large fields are very similar in both structural phases
(and the free molecule), they are not important for understanding
relative phase stability. Thus, changes in the pressure stress field
highlight regions and charge density features that contribute to
favorable energetic changes. 

Figure \ref{fig2} contrasts the pressure fields for $\beta=0$
and $\beta=1/8$. 
Note that there is no qualitative change in the pressure field within 
the interstitial regions since the charge density in this area 
is quite smooth and hence the value of its Laplacian is small 
(averaging 0.05 ~eV/\AA$^3$). 
Therefore our above analysis holds regardless of our choice of $\beta$. 

\section{Conclusion}
We have developed a new formulation of the quantum stress field
using differential geometry. We identify the stress tensor field 
as a variation of the total ground state quantum-mechanical energy 
with respect to the metric tensor field.
Within this formulation we have uniquely specified the 
stress field up to a single ambiguity. 
The origin of the ambiguity lies in the inability to uniquely 
describe the dependence of the kinetic energy on the metric
tensor. 

We have successfully extended this formulation to
local density functional theory. 
Terms identical to the Maxwell stress can be found within the 
LDA-DFT stress field. We note that within the gravitational 
literature, it is well-known that the Maxwell stress is unique.
We describe the similarities between the pressure field obtained
by our method and the various terms found in the total energy
density. We then show that the pressure field contains an ambiguity
in the form of the Laplacian of the charge density multiplied by a 
constant $\beta$. Through the relationship between terms in the pressure field
and the kinetic energy density  we offer a realistic range of choices
for the value of $\beta$. 

To demonstrate the usefulness of stress field
as a qualitative tool in combination with total energy calculations, 
we used the pressure field to rationalize the energetics of two phases of 
solid molecular hydrogen. We finally show that our conclusions remain
invariant with respect to realistic choices of $\beta$.

\section{Acknowledgements}
The authors wish to acknowledge V. Balasubramanian, J. Bassani, K. Burke, R. P. Kauffman, R. M. Martin,  
P. Nelson, and D. Vanderbilt for their comments and valuable
suggestions regarding this work, and E. J.  Walter for his assistance
with the numerical calculations.  This work was supported by NSF grant
DMR 97-02514 and the Air Force Office of Scientific Research, Air
Force Materiel Command, USAF, under grant number F49620-00-1-0170. AMR
would like to thank the Camille and Henry Dreyfus Foundation for
support. Computational support was provided by the National Center for
Supercomputing Applications and NAVOCEANO MSRC.

\begin{figure}

\vspace{0.5cm}
\caption{
Contour plots of the pressure field with $\beta=0$ and the charge density within DFT for
the \textit{Cmca} structure (panels A and C) and for the \textit{m}-hcp structure (B and
D) of solid hydrogen.  The vertical axis is the stacking direction for
the layers, and the horizontal axis is the direction along which
alternating layers are shifted.  The plots are 6.794 \AA ~high and 7.206 \AA 
~wide. Ten contours are shown, over a range of 0--3~eV/\AA$^3$ for the
pressure fields, and from 0--2.5~\textit{e}/\AA$^3$ for the charge densities.
}
\label{fig1}
\end{figure}

\begin{figure}

\vspace{0.5cm}
\caption{
Contour plots of the pressure field in solid hydrogen for the \textit{Cmca} and \textit{m}-hcp structures  with $\beta=0$
(panels A and C), and $\beta=1/8$ (panels B and D).
The vertical axis is the stacking direction for
the layers, and the horizontal axis is the direction along which
alternating layers are shifted.  The plots are 6.794 \AA ~high and 7.206 \AA 
~wide. Ten contours are shown, over a range of 0--3~eV/\AA$^3$.
}
\label{fig2}
\end{figure}


\begin{references}
\bibitem{units} Choosing $c$ dimensionless is equivalent to measuring distance and time with the same units, giving energy and mass the same
units. Making $\hbar$ dimensionless gives energy (and mass) the units of inverse time (or inverse length).
\bibitem{calc2} Calculations for both structures were performed in an orthorhombic unit cell with parameters $a=2.08$, $b$, and $c$ with $b/a=\sqrt{3}$ and
$c/a=\sqrt{8/3}$. All atomic forces were relaxed to $<$ 5 meV/\AA.
\end{references}
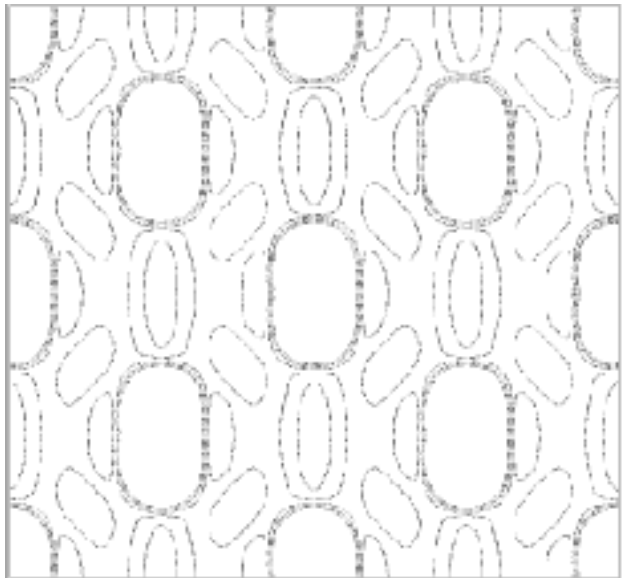
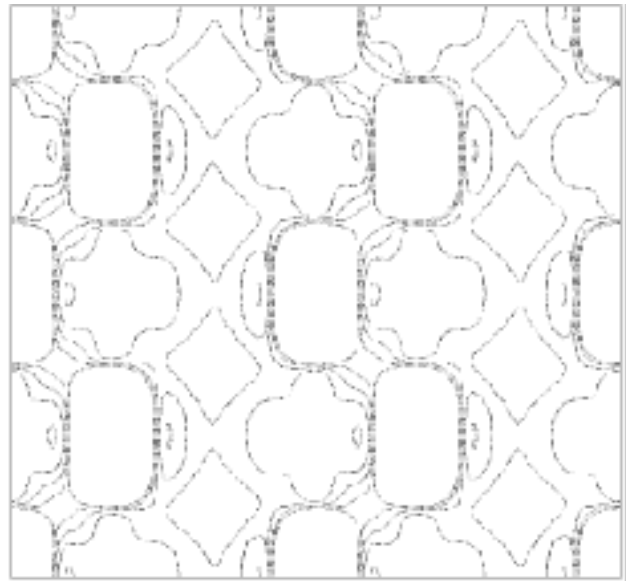
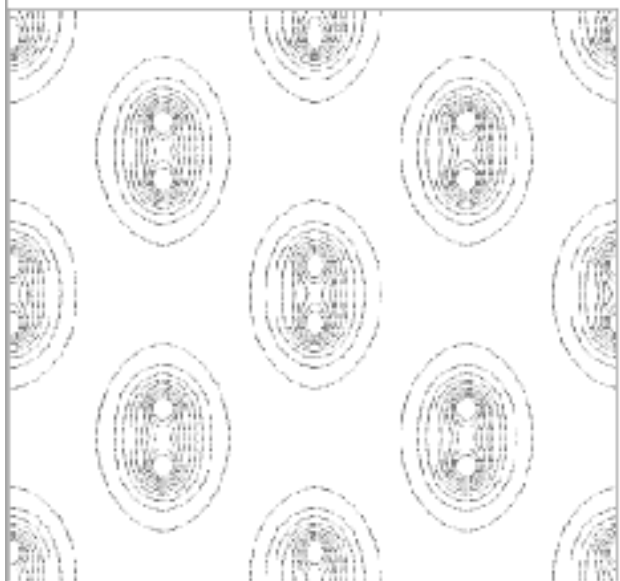
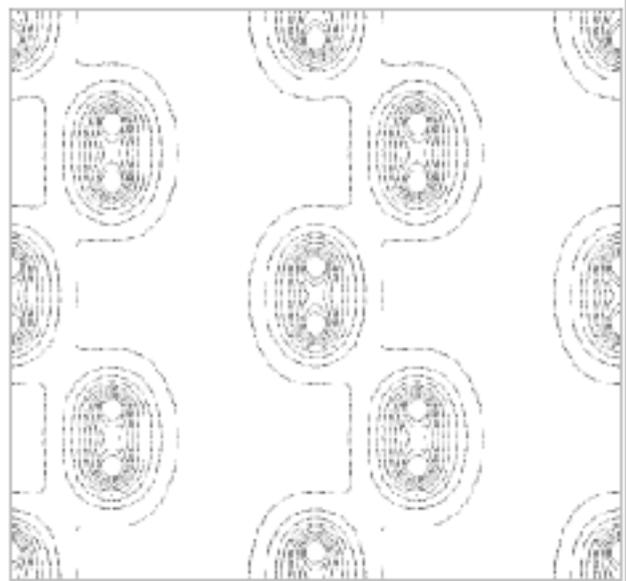
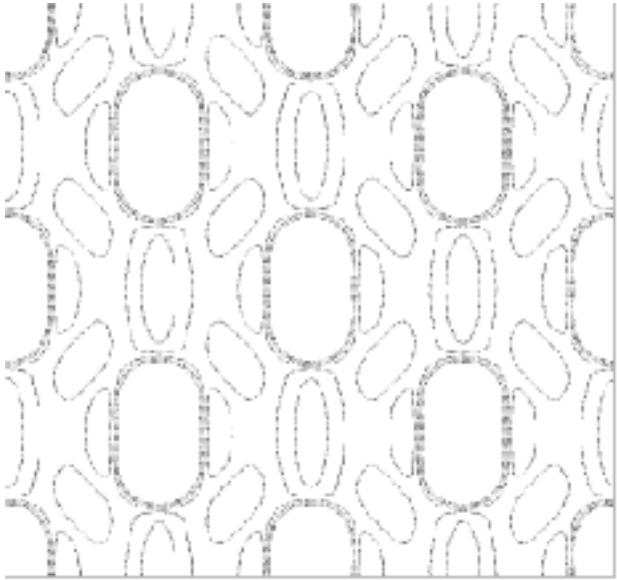
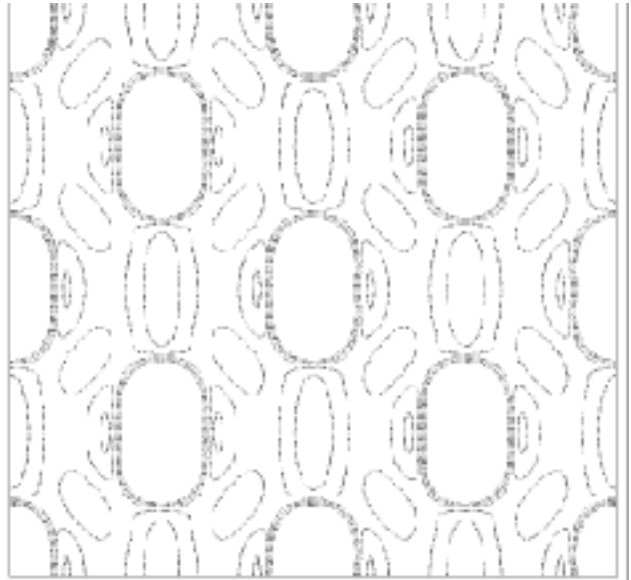
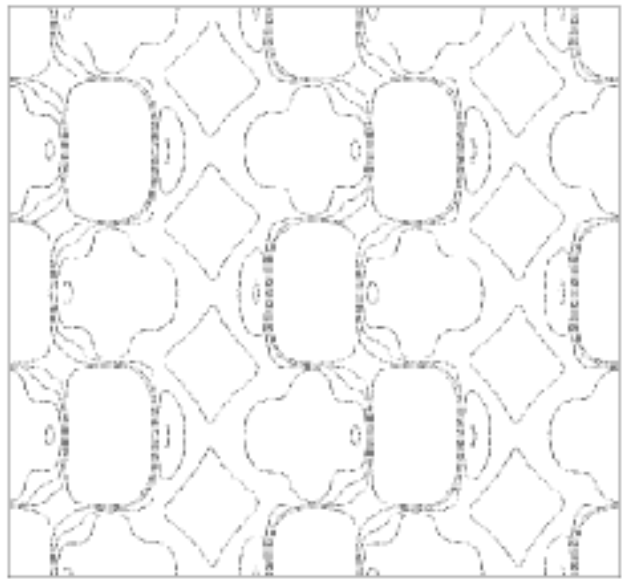
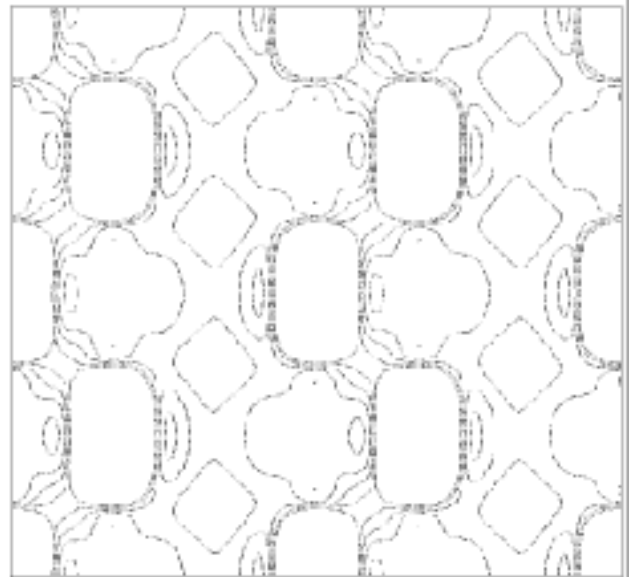
\end{document}